\documentstyle[epsf]{mn}

\def\simlt{\mathrel{\rlap{\lower 3pt\hbox{$\sim$}}
        \raise 2.0pt\hbox{$<$}}}
\def\simgt{\mathrel{\rlap{\lower 3pt\hbox{$\sim$}}
        \raise 2.0pt\hbox{$>$}}}

\title[The Observed Evolution of Galaxy Clustering vs.
Epoch-Dependent Biasing Models]
       {The Observed Evolution of Galaxy Clustering vs.
       Epoch-Dependent Biasing Models}

\author[M.Magliocchetti, J.S.Bagla, S.J.Maddox, O. Lahav]
       {M.Magliocchetti $^1$, J.S.Bagla $^{1,2}$,
        S.J.Maddox$^{1,3}$, O.Lahav$^{1,4}$ \\
        $^1$Institute of Astronomy, Madingley Road, Cambridge CB3 0HA\\
        $^2$Harvard-Smithsonian Center for Astrophysics, Mail Stop 51,
        60 Garden Street, Cambridge, MA 02138, U.S.A.\\
        $^3$School of Physics and Astronomy, University of Nottingham,
        Nottingham, NG7 2RD, UK.\\
        $^4$Racah Institute of Physics, The Hebrew University,
        Jerusalem 91904, Israel}

\pagerange{\pageref{firstpage}--\pageref{lastpage}}
\pubyear{1999}

\begin{document}
\label{firstpage}

\maketitle

\begin{abstract}
We study the observed evolution of galaxy clustering as a function of
redshift.  We find that the clustering of galaxies, parameterized by
the amplitude of fluctuations in the distribution of galaxies at a
co-moving scale of 8$h^{-1}$~Mpc, decreases as we go from observations
of the local Universe to $z \sim 2$.  On the other hand, clustering
of the Lyman break galaxies at $z \sim 3$ is very strong, comparable
to the clustering of present day galaxies. 

However there are three major factors to take into account while
comparing clustering measurements coming from various surveys: the
so-called ``scale-dependence'' effect, due to measurements being made
at different scales; the ``type-selection'' effect introduced by the fact
that different galaxy surveys select different populations which do
not have the same clustering amplitudes; and the Malmquist bias which
means that within a given survey the more distant galaxies tend to
have brighter absolute magnitudes, and so do not have the same
clustering amplitude.
We correct for the first two effects and discuss the implications of
Malmquist bias on the interpretation of the data at different
$z$. Then we compare the observed galaxy
clustering with models for the evolution of clustering in some fixed
cosmologies.
Correcting for the scale-dependence effect significantly reduces the 
discrepancies amongst different measurements. 

We interpret the observed clustering signal at high redshift as coming
from objects which are highly biased with respect to the underlying
distribution of mass; this is not the case for $z\simlt 2$ where
measurements are compatible with the assumption of a much lower
biasing level which only shows a weak dependence on $z$.
Present observations still do not provide a strong constraint because
of the large uncertainties but clear distinctions will be possible
when larger datasets from surveys in progress become
available. Finally we propose a model-independent test that can be
used to place a lower limit on the density parameter $\Omega_0$.

\end{abstract}


\section{INTRODUCTION}

It is believed that structures like galaxies and clusters of galaxies
formed by accretion of matter onto small inhomogeneities present in
the early Universe.  
The simplest models assume that the distribution of galaxies is
directly related to the underlying density distribution and the two
distributions evolve in a similar manner.
This has provided a key motivation for redshift surveys of galaxies. 
However, many studies have shown that the relation between galaxy
clustering and that of underlying matter is not simple and this
relation is, in general, a function of time
\cite{tgb_corr,Fry,Mo,Bag,DL,teg,Narayanan,bias_phy,pedro,durham,garching}.
These studies deal with the evolution of the clustering of dark matter
halos and many factors, including observational selection functions
and evolution of stellar populations in galaxies have to be taken into
account before these results can be applied to real observations.  Some
comparisons of models and observations have been carried out
\cite{Mat,Mos} where a large class of models were compared with the
available observations.

Models of halo clustering \cite{Mo,Bag} suggest that more massive
halos cluster more strongly than halos of lower mass.  This follows
from the simple model for biasing \cite{kaiser} in which rare objects
cluster more strongly than the more typical objects.  If, as
observations suggest, the luminosity of a galaxy increases
monotonically with the mass of the halo in which it resides, then we
expect brighter galaxies to cluster more strongly than fainter
galaxies.
The same argument also suggests that amongst halos of a given mass,
older halos cluster more strongly.  If the gravitational clock is
synchronised with the stellar clock then we expect early type galaxies
to cluster more strongly than late type galaxies.  This is seen in
simulations \cite{bias_phy} that include simple recipes for star
formation.

A common conclusion of all theoretical and numerical studies of halo
clustering is that the rate of evolution of halo clustering,
$D_{halo}(t)$, is always slower than the rate of evolution of
clustering in dark matter $D_{m}(t)$, so that $\dot{D}_{halo}(t) \leq
\dot{D}_{m}(t)$, where the dot represents differentiation with respect
to time.  These rates are equal only in the limit when all the matter
has collapsed into halos.

Turning now to a brief discussion of effects that influence the
observed evolution of galaxy clustering, consider a Universe in which
the galaxies do not evolve: neither in their stellar content, nor in
their distribution in space.  So the clustering is fixed in comoving
space and galaxies at all redshifts are similar to the ones we see in
the local Universe.  What will be the observed amplitude of clustering
at different redshifts, if we conduct a magnitude limited redshift
survey in such a Universe?  As nothing is changing as far as galaxies
are concerned, the only differences are given by observational
selection effects.  There are at least two of these: Malmquist bias
and K-correction.  Let us examine the effect of these two factors
separately.

In an apparent magnitude limited survey, we will only see brighter
galaxies at high redshifts whereas at lower redshifts we will also see
fainter galaxies.  Since brighter galaxies tend to cluster more
strongly than fainter galaxies \cite{park_cfa2,Lo}, the effect of
Malmquist bias in our imaginary survey will lead to an apparent
increase in the amplitude of clustering with redshift.  Also in an
apparent magnitude limited sample, the observed clustering amplitude
will always exceed the true clustering amplitude of all galaxies at
that redshift.  This variation with redshift can be avoided  by using
a cut in absolute magnitude instead of apparent magnitude, but the
inferred absolute magnitudes depend explicitly on the assumed values
of cosmological parameters.  However, without going into any details
of dependence on cosmological parameters and models of galaxy
evolution, we can conclude that the observed rate of evolution of
clustering amplitude is always smaller than the true rate in a
magnitude limited survey, the only assumption here being that brighter
galaxies cluster more strongly than fainter
galaxies:~$\dot{D}_{obs}(t) \leq \dot{D}_{true}(t)$. 

The effect of K-correction depends strongly on the wave-band used to
define the sample. For example, the difference between the rest-frame
B~band luminosity for ellipticals and irregulars observed in the 
B~band is $\sim 4$ magnitudes at $z=1$, but is only $\sim 2$
magnitudes if they are observed in the I~band  (Tresse 1999).
This effect means that the relative populations of early- and late-type
galaxies in a B~band selected survey will vary strongly as a function
of redshift, preferentially selecting late-types at higher redshift.
Since later-type galaxies cluster less strongly than early types
(e.g. Loveday et al. 1995; Hermit et al. 1996; Guzzo et al. 1997;
Loveday Tresse \& Maddox 1999) a sample defined in the optical will
generally tend to underestimate the correlation amplitude at higher
redshifts.
In the near infra-red the spectral energy distributions of different
galaxy types are very similar to each other, and so the K-corrections
are very similar. 
Hence the problem is less significant if we use near infra-red
wavelengths to define the sample.
Quantifying this effect requires knowledge of the relative fractions
of different galaxy populations as a function of redshift, and of how
the clustering varies for each population. We plan to tackle this
problem in a future paper, but in our present analysis we have simply
grouped the various measurements according to the survey selection
criteria so that the differences between types are minimized.

This problem could be by-passed if a fixed rest-frame bandpass were 
used to define the sample, but even this would not take into
account the spectral evolution of galaxies.  
In the real Universe, all galaxies tend to get bluer at higher redshifts, 
because of stellar evolution, and so the differences between different 
types of galaxies become smaller.  
This means that in the real Universe, uncertainties due to different
K-corrections will be smaller than in our imaginary Universe.

Another factor which affects the interpretation of the data from
various surveys is the ``scale-dependence'' of the 
measurements: different surveys sample clustering on different
physical scales.
Even though the values for the correlation length $r_0(z)$ may be of
the order of a few Mpc's, many surveys (especially those at high
redshifts) do not sample scales above $\sim$1 h$^{-1}$ Mpc.
Since the correlation function is not necessarily a pure power law and
the measurements have different slopes, this effect can introduce
systematic offsets between the various surveys.  To take this into
account in this paper, we always  compare theoretical predictions to
data at the same scale.

This paper presents a compilation of clustering measurements and aims
to introduce some simple scenarios for the evolution of clustering in
order to understand the different factors which enter into the
comparison of data and models.
More detailed quantitative discussion on the evolution of clustering
will require corrections for all of the effects discussed above,
possibly through comparison with a library of models for galaxy formation
(e.g. semi-analytical models) convolved with observational selection
effects. 

The layout of the paper is as follows: section 2 introduces simple
analytical models for the evolution of bias with epoch, while section
3 presents the observations and shows the trend of the ``raw'' data
(i.e. with no corrections applied for the effects described in this
section) together with simple models for the evolution of
clustering. A more detailed analysis of the data, together with our
interpretation of the results is given in section 4. Section 5
introduces a model-independent test to place a lower-limit on the
density parameter $\Omega_0$, while section 6 summarises our
conclusions.

\section{Evolution of Bias}

Most theoretical models for evolution of galaxy clustering identify
galaxies with halos. Here we will describe a few models for the
redshift evolution of the bias of halos $b(z)$.

\begin{itemize}
\item {\bf No Evolution}~(B0) : In this model bias does not change
with redshift and remains constant at its present value.  This
assumption does not have any physical basis and this model serves only 
as a reference.  The bias is defined as
\begin{equation}
b_0 = \frac{\sigma_{8,g}}{\sigma_{8,m}} ,
\label{eqn:b0}
\end{equation}
where $\sigma_8$ is the \rm{r.m.s.} density fluctuation at 8$h^{-1}$
Mpc, subscripts $g$ and $m$ denote galaxies and total underlying mass,
respectively and all quantities are evaluated at $z=0$.\\
As the halo correlation function evolves at a slower rate than the
dark matter one, we expect bias to be higher at high redshifts than at
present. Therefore this model should underestimate $\sigma_{8,g}(z)$
at high $z$.

\item {\bf Test Particle Bias}~(B1) : This model does not assume
anything about the origin of halos, or about their initial distribution.
Test particles are distributed through the Universe such that
their density contrast is proportional to the density contrast of the
total underlying mass. The model describes the evolution of
bias for these test particles by assuming that they follow the cosmic
flow.  It can be shown \cite{Nu,Fry,teg} that the bias for such test
particles evolves as
\begin{equation}
b(z) = 1 + \frac{b_0 - 1}{D(z)},
\label{eqn:b1}
\end{equation}
where $D(z)$ is the linear growth rate for clustering.
Here $b_0$ is the bias for the set of halos/test particles at present
epoch (see equation \ref{eqn:b0}). This bias does not depend on the
mass of halos and the model
works well in the range $0 \simlt z \simlt 1$ for CDM like models
\cite{Bag} if the halo distribution is biased. Note that the predicted
variation for anti-biased halos ($b_0 <1$) is not seen in simulations simply
because the basic premise of inert, indestructible halos is not
correct.  This model is also called the {\it galaxy conserving model}
\cite{Mat}.

\item {\bf Merging Model}~(B2) : This model \cite{Mat} for the evolution
of galaxy bias  allows halos to undergo dissipative collapse
(i.e. merging). 
It is based on the Mo and White~(1996) model for halo bias which was
in turn computed, from the formalism of Press and Schechter (1974).
The  bias is computed for all halos with mass $M$ above a certain
threshold $M_{min}$ which  is computed by normalizing the effective
bias $b^{\it eff}$ at $z=0$ to the observed bias.  
The generic expression for bias of halos of mass $M>M_{min}$ is given by
\begin{equation}
b(M,z)=b_{-1}+(b_0^{\it eff}-b_{-1})/D(z)^{\beta},
\label{eqn:b2}
\end{equation}
where the parameters $b_0^{\it eff}$ and $\beta$ depend on the choice of
$M_{min}$ and the background cosmology.  As in the model of Mo and
White~(1996), there is also some dependence on the formation redshift
of halos, which in this case was fixed to a generic $z_F$. The
constant value $b_{-1}=0.41$ is obtained in the limit
$M_{min}\rightarrow 0$. 

In the present study, we decided to leave $M_{min}$ as a free
parameter (ranging $10^{10}M_{\odot} \le M_{min} \le
10^{13}M_{\odot}$) so as to allow the objects under consideration to
be either transient or to have changed their properties with time
(c.f. {\it transient model} \cite{Mat}), and to keep the discussion on
a more general level.
\end{itemize}
As already stated, both the values of $b_0^{\it eff}$ and $\beta$ and 
those for $b_0$ (and therefore of $\sigma_{8,m}$ through equation
\ref{eqn:b0}) are determined by the choice of the cosmological
parameters and power spectrum of mass fluctuations. 
In the following models we will consider three combinations, namely
\begin{itemize}

\item I :
$\Omega_0=1\;\; \Omega_{\Lambda,0}=0\;\ h_0=1\;\; \Gamma=0.25$\\
$\sigma_{8,m}^{\rm{lin}}=0.65$

\item II :
$\Omega_0=0.4\;\; \Omega_{\Lambda,0}=0\;\ h_0=0.65\;\; \Gamma=0.23$ \\
$\sigma_{8,m}^{\rm{lin}}=0.64$

\item III :
$\Omega_0=0.4\;\; \Omega_{\Lambda,0}=0.6\;\ h_0=0.65\;\; \Gamma=0.23$ \\
$\sigma_{8,m}^{\rm{lin}}=1.07$

\end{itemize}
The normalization of the linear \rm{r.m.s.} mass density fluctuations
at 8$h^{-1}$ Mpc $\sigma_{8,m}^{\rm{lin}}$ for each power spectrum is chosen to
match the four year COBE DMR observations \cite{Bunn}. The values of the
parameters $b_0^{\it eff}$ and $\beta$ in equation (\ref{eqn:b2}) as
functions of different cosmologies and different halo masses are given
in Matarrese et al.~(1997) and Moscardini et al.(1998).  However,
much of the discussion is independent of the specific normalization or
the shape of the dark matter power spectrum.\\

\noindent
We note that biasing is likely to be non-linear, non-local,
scale-dependent, type-dependent and stochastic (e.g. Dekel \& Lahav
1998; Tegmark \& Peebles 1998; Blanton et al. 1998; Narayanan, Berlind
and Weinberg 1998), so the models discussed here will give a  highly
simplified picture of galaxy clustering evolution. 

\section{Observations}
\begin{figure*}
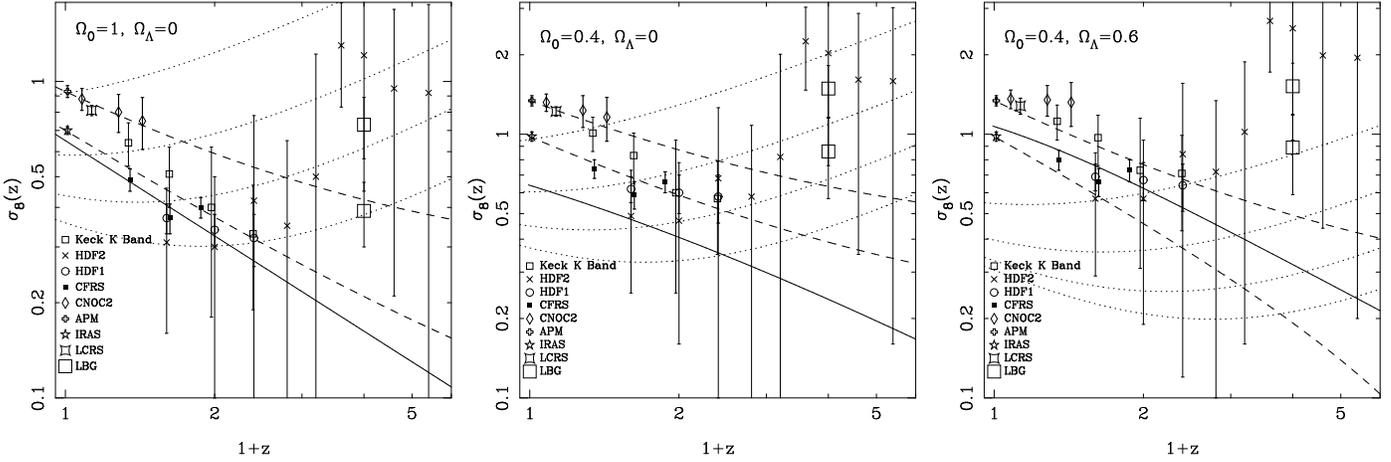

\vspace{8cm}  
\includegraphics{jasEdS.ps}
\includegraphics{jasop.ps}
\includegraphics{jaslam.ps}
\caption{Measurements of $\sigma_8$ as a function of redshift
  respectively for an Einstein-deSitter Universe $\Omega_0=1$, $h_0=1$
  (left panel), OCDM $\Omega_0=0.4$, $h_0=0.65$ (middle panel) and
  $\Lambda$CDM $\Omega_0=0.4$, $\Omega_{\Lambda,0}$, $h_0=0.65$ 
(right panel). The
error bars are the quoted $68\%$ confidence limits or the equivalent
taken from the papers cited in the description of surveys used here.
The thick line shows the linear rate of growth for dark matter.  The
dashed curves show evolution as predicted by the test particle
model, the two curves are anchored to the APM and the IRAS
observations at $z=0$.  The dotted curves show the evolution of bias in
the merging model for $10^{10}$, $10^{11}$, $10^{12}$ 
and $10^{13}$M$_\odot$ from bottom upwards.}
\label{fig:fit1}
\end{figure*}
In this section we  put together the observations of galaxy
clustering from different surveys.
In the following we assume that the correlation function has a
power-law form $\xi(r,z)=\left[r/r_0(z)\right]^{-\gamma}$ at the
relevant scales, and hence the index $\gamma$ and the correlation
length $r_0$ are sufficient to describe it at a given redshift.  Most
observers quote these numbers and in cases where these were given in
proper coordinates, we converted them to the corresponding comoving
scales.  To transform results between different cosmologies, we use
the following expression
\begin{equation}
r_{02}(z)=\left[\frac{h_{01}}{h_{02}}\left(\frac{x_1(z)}{x_2(z)}
\right)^{1-\gamma}\frac{P(\Omega_{01},z)}{P(\Omega_{02},z)}\;
\frac{F_1(z)}{F_2(z)}\right]^{1/\gamma} r_{01}(z)  .
\label{eqn:rr}
\end{equation}
This equation is derived by requiring that the angular correlation
function of a set of galaxies between redshift $z$ and $z + \Delta z$
is the same in different cosmologies.  If $\Delta z$ is small enough
for us to assume a constant redshift distribution of objects $N(z)$ 
then the relativistic Limber
equation \cite{Pe} leads us to equation \ref{eqn:rr} for a power law
correlation function.  The relative expressions for the comoving
coordinate $x$ and the functions $P$ and $F$ for different 
geometries are given in Appendix 1 (see also Magliocchetti and Maddox,
1999, Treyer and Lahav, 1996).

To compare different datasets, we will use the \rm{r.m.s.}
fluctuations in the galaxy distribution at the scale of
8$h^{-1}$~Mpc, $\sigma_{8}$.

We relate $\sigma_8$ to the other two parameters as
\cite{Pe}
\begin{eqnarray}
\sigma_8(\bar z)=\left[\left(\frac{r_0(\bar z)}{8}\right)^
{\gamma}c_{\gamma}\right]^{1/2},
\label {eqn:sigma} \\
c_{\gamma}=\frac{72}{(3-\gamma)(4-\gamma)(6-\gamma)2^{\gamma}}.
\end{eqnarray}

The list of surveys from which the data points have been taken is
given below and in table~1, along with detailed information about the relevant
parameters of each survey. These parameters include the median
redshift $\langle z \rangle$ of the survey, the number of galaxies
$N_{GAL}$ in each survey, the angular coverage, the
selection band, the range of absolute magnitudes L (if available) of the
objects included in the analysis of clustering and the relative 
values of $\gamma$. We considered the following surveys:
\begin{table*}
\caption{Summary of the properties of each survey.}
\begin{flushleft}
\begin{tabular}{lllllllllll}
$Survey$& $\langle z \rangle$& $N_{GAL}$ & Angular Coverage & Selection band &
Absolute Magnitude&$\bar
z $ & $\gamma$\\
\hline
IRAS & $\le 0.1$ & 9080&$|b|>10^{\circ}$&$F_{60}\le 0.6Jy$&?& $\sim
0.06$ & $1.57\pm 0.03$\\
\small{Stromlo-APM} &0.06 & 1757&4300 deg$^2$&$b_j\le
17.5$&$-22\le L_B\le -15$ &$\sim 0.05$ & $1.71\pm 0.05$\\
\small{Las Campanas}&0.1& 19558&700 deg$^2$&$R\le 17.75$ &$-23\le L_R\le -18$ &
$0.13$ & $ 1.85\pm 0.05$\\
CNOC2&0.36 & $\sim 2000$ &$40\times9'\times
8'$&$R\le 24$& $L_R\ge -20$ & $0.08$ &$ 1.8\pm 0.1$\\
CNOC2 &0.36 & $\sim 2000$ &$40\times9'\times
8'$&$R\le 24$& $L_R\ge -20$ &$0.28$ & $1.8\pm 0.1$\\
CNOC2 &0.36 & $\sim 2000$  &$40\times 9'\times
8'$&$R\le 24$& $L_R\ge -20$ &$0.43$ &$1.8\pm 0.2$\\
CFRS  &0.56 & 591&$5\times10'\times 10'$&$17.5\le I_{AB}\le
22.5$&$-21.5\simlt L_B\simlt -18.5$ &$0.35$ &$1.64\pm 0.05$\\
CFRS &0.56 & 591&$5\times10'\times 10'$&$17.5\le I_{AB}\le
22.5$& $-22.0\simlt L_B\simlt -19.5$ &$0.62$ & $1.64\pm 0.05$\\
CFRS &0.56 & 591&$5\times10'\times 10'$&$17.5\le I_{AB}\le
22.5$&$-22.5\simlt L_B\simlt -21.0$ &$0.87$ 
&$1.64\pm 0.05$\\
\small {Keck K-band}  &0.7 & 248&27 arcmin$^2$&$K\le 20$  
&$L_K\ge -21.5$&$0.34$&  1.8\\
\small {Keck K-band} &0.7 & 248&27 arcmin$^2$&$K\le 20$ 
&$L_K\ge -23.5$& $0.62$  &1.8\\
\small {Keck K-band}&0.7 & 248&27 arcmin$^2$&$K\le 20$
&$L_K\ge -23.5$& $0.97$ &1.8\\
\small {Keck K-band}&0.7 & 248&27 arcmin$^2$&$K\le 20$ 
&$L_K\ge -23.5$& $1.39$ &1.8\\
HDF$_1$  &$ 1$ & 926&5 arcmin$^2$&$I_{AB}\le 27$ &?&$0.6$ & 1.8\\
HDF$_1$  &$ 1$ & 926&5 arcmin$^2$&$I_{AB}\le 27$ &?&$1.0$ &1.8\\
HDF$_1$  &$ 1$ & 926&5 arcmin$^2$&$I_{AB}\le 27$&? &$1.4$ &1.8\\
HDF$_2$  &$ 1.6$ & 946&4 arcmin$^2$&$AB(8140)\le 28$& ?
&$0.6$ &1.8\\
HDF$_2$ &$ 1.6$ & 946&4 arcmin$^2$&$AB(8140)\le 28$ &?& $1.0$&1.8\\
HDF$_2$ &$ 1.6$ & 946&4 arcmin$^2$&$AB(8140)\le 28$ &?& $1.4$ &1.8\\
HDF$_2$ &$ 1.6$ & 946&4 arcmin$^2$&$AB(8140)\le 28$&? &$1.8$&1.8\\
HDF$_2$&$ 1.6$ & 946&4 arcmin$^2$&$AB(8140)\le 28$ &? &$2.2$&1.8\\
HDF$_2$ &$ 1.6$ & 946&4 arcmin$^2$&$AB(8140)\le 28$ &? &$2.6$  &1.8\\
HDF$_2$ &$ 1.6$ & 946&4 arcmin$^2$&$AB(8140)\le 28$ &? &$3.0$ &1.8\\
HDF$_2$ &$ 1.6$ & 946&4 arcmin$^2$&$AB(8140)\le 28$ &? &$3.6$&1.8\\
HDF$_2$&$ 1.6$ & 946&4 arcmin$^2$&$AB(8140)\le 28$ &? &$4.4$ &1.8\\
LBG$_1$ &3 & 871& $9\times\sim 9'\times 9'$&$R\le 25.5$ &? &$3.0$ 
&$1.98\pm 0.3$\\
LBG$_2$&3 & 268&$6\times\sim 9'\times 9'$&$R\le 25.5$ &? &$3.0$ &1.8\\
\end{tabular}
\end{flushleft}
\end{table*}
\begin{itemize}

\item Stromlo-APM  \cite{Lo}.

\item IRAS  \cite{Sa}.

\item Las Campanas  \cite{lascamp}.

\item CFRS \cite{Le}.

\item HDF$_1$ (Connolly, Szalay \& Brummer,1998).

\item HDF$_2$ (Magliocchetti \& Maddox, 1999).

\item Keck K-band \cite{Car}.

\item CNOC2 \cite{Car1}.

\item LBG$_1$ (Giavalisco et al., 1998).

\item LBG$_2$ (Adelberger et al., 1998).

\end{itemize}
The values for $\sigma_8(\bar{z})$ for the three cosmologies presented
in section 2 are shown in Figure~1. 
Note that there are two different points 
for Lyman break galaxies (LBG) at
$z=3$.  These correspond to two different subsamples -- one with
observed redshifts that can generally be described as the brighter
sample, and the other with photometric redshifts.  In the first case
the amplitude of fluctuations is determined by counts-in-cells
\cite{Ad} and in the other case it is arrived at through the angular
correlation function \cite{Gia}.  The higher amplitude point corresponds to the
sample with redshifts.\\
In order to guide the eye we also plotted some simple models for the
evolution of clustering. Here we assume $\sigma_{8,m}(z)$ to vary with
redshift according to linear theory, i.e. we write 
\begin{eqnarray}
\sigma_{8,g}(z) = b(z)\; \sigma_{8,m}^{\rm lin}(z),\;\;\rm{with}\;\;\;\;\;\;\nonumber \\
\sigma_{8,m}(z)=\sigma_{8,m}^{\rm lin}(z=0) \frac{D(z)}{D(z=0)},
\end{eqnarray}
($D(z=0)=1$ by definition), and $b(z)$ given by equations
(\ref{eqn:b0}-\ref{eqn:b2}). 
The thick line shows the linear rate of
growth for dark matter.  It is clear that within any given survey, the
amplitude of fluctuations does not fall as rapidly as the linear
rate.  This is encouraging because according to the arguments outlined
in the introduction, the observed rate of evolution should  always be smaller
than the  rate of evolution of mass.  The dashed curves show evolution
as predicted by the test particle model (B1), the two curves are anchored
to the APM and the IRAS observations at $z=0$.  The dotted curves show
the evolution of bias in the merging model (B2) for $10^{10}$, 
$10^{11}$, $10^{12}$ and $10^{13}$M$_\odot$ from bottom upwards.\\
The basic pattern followed by the amplitude of clustering, 
even though it is masked to some extent by large error bars and 
differences in different datasets, 
is that - independent of the cosmological model - 
at low redshifts, $\sigma_{8,g}$ decreases
with increasing redshift, reaches a minima around $z=2$ and then rises
again at higher redshifts.  This type of variation has been seen for
dark matter halos in N-body simulations (see e.g Jenkins et al. 1998),
but given the observational complications discussed earlier, these simulations 
cannot be directly compared to the observational data.

\section{Interpretation}

Given the general trend shown in Figure 1, we have to address the following
questions:~(1)~Why does the observed clustering vary across datasets?
(2)~Is it possible to scale different datasets to make a
self-consistent dataset, and then study evolution within that superset?
(3)~Can we constrain any of the models using this data?

There are significant discrepancies between the amplitude of clustering
in different surveys even where they sample the same redshift intervals.
To some extent these discrepancies are intrinsic because different surveys 
sample different populations of galaxies.  Some differences are introduced 
by the extent to which the luminosity function is probed, i.e. in one 
survey the limiting magnitude may allow one to probe galaxies 
much fainter than L$^*$ and in another case the limiting 
magnitude may be comparable to L$^*$.  
Since different surveys measure clustering at different scales, and
many of them extrapolate the values for $r_0(z)$ on scales bigger than
the areas actually covered by the surveys themselves (see table 1),
further differences are  introduced if the galaxy correlation
function is not a true power law in the range of scales between the
scale of measurement and 8$h^{-1}$Mpc.

All these effects are very likely to ``bias'' the different
measurements with respect to each other. In this section we 
tackle these issues one by one, in order to correct the data for their
effects so to obtain ``unbiased'' sub-sets of measurements
compatible with each other.

\subsection{Scale Dependence}

As already mentioned in the previous sections and as shown in
table 1, different surveys cover areas of the sky which
vary greatly in size from one survey to another. This problem of sampling
objects on different scales is likely to introduce a relative bias
amongst clustering measurements coming from different surveys. In fact
many of these studies quote values for the clustering length $r_0(z)$
which have been obtained by extrapolating the power-law 
trend of $\xi(r,z)$ to scales
much larger than the physical scales of the surveys. In
order to correct for this effect we write the bias as
\begin{eqnarray}
b^2(\bar{r},z)=\frac{\xi_g(\bar{r},z)}{\xi_m(\bar{r},z)}, 
\end {eqnarray}
where $\xi_m$ and $\xi_g$ respectively are the mass-mass and
galaxy-galaxy correlation function and $\bar{r}$ is some fiducial
scale length.
Note that by writing the bias as an explicit function of $\bar r$
which is different for each survey, this approach corrects for the
``scale effect'', since it compares theoretical quantities and
measurements evaluated at the same scale. 

We start by evaluating the mass-mass correlation function 
$\xi_m(\bar{r},z)$. 
Throughout this section we will use the notations adopted in
Peacock (1997). We start with a (dimensionless) 
primordial power-spectrum of the 
form $\Delta_{lin}^2(k)\propto k^{n+3}T_k^2$ (with $n=1$ for CDM models). 
The transfer function for CDM family of models is 
the one given by Bardeen et al. (1986):
\begin{eqnarray}
T(k)=\frac{\rm{ln}(1+2.34q)}{2.34q}\times
\;\;\;\;\;\;\;\;\;\;\;\;\;\;\;\;\;\;
\nonumber\\
\left[1+3.89q+(16.1q)^2+(5.46q)^3+(6.71q)^4\right]^{-1/4},
\end{eqnarray} 
where $q=(k/h)/\Gamma$ Mpc$^{-1}$ and the shape parameter $\Gamma$ is
related to the present-day matter parameter $\Omega_0$ and the
baryonic fraction $\Omega_{0b}$ via
$\Gamma=\Omega_0\; h\;\rm{exp}\;[\Omega_{0b}-\sqrt{h/0.5}\;\Omega_{0b}/\Omega_0]$
(Sugiyama, 1995).
The normalization of the power-spectrum is fixed by specifying the
values of $\sigma_8^{\rm{lin}}$ (see section 2), according to the expression:
\begin{eqnarray}
(\sigma_R^{\rm{lin}})^2=\int\Delta_{\rm{lin}}^2(k)\frac{dk}{k}\frac{9}{(kR)^6}
\left[\sin kR-kR \cos kR\right]^2
\end{eqnarray}

In order to go from the linear power-spectrum 
$\Delta_{\rm{lin}}^2(k)$ to the non-linear case, Peacock \& Dodds
(1994, 1996), following an 
approach originally introduced by Hamilton et al. (1991), assumed there is a
universal fitting function F relating the two according to the expressions:
\begin{eqnarray}
\Delta^2(k)=\rm{F}\left[\Delta_{lin}^2(k_0)\right], \nonumber\\
k_0=\left[1+\Delta^2(k)\right]^{-1/3}k,
\end{eqnarray}
where $k_0$ and $k$ are respectively the linear and non-linear
wavenumber and F is given by:
\begin{eqnarray}
F(x)=x\left[\frac{1+B\beta x+[Ax]^{\alpha
\beta}}{1+([Ax]^{\alpha}g(\Omega_0)^3/[Vx^{1/2}] )^{\beta}}\right]^{1/\beta}
\label{eq:F}
\end{eqnarray}
(see Peacock
\& Dodds (1996) for the values of the parameters in equation
\ref{eq:F}). $g(\Omega_0)$ is a suppression factor which measures the rate of
growth of clustering in generic cosmologies relative to the growth in
an Einstein-de Sitter Universe.  Lahav et al. (1991) and 
Carrol, Press \& Turner (1992) found that 
this quantity can be almost exactly approximated by
\begin{eqnarray}
g(\Omega_0,\Omega_{\Lambda,0})=
\frac{5}{2}\Omega_0\left[\Omega_0^{4/7}-\Omega_{\Lambda,0}+(1+\Omega_0/2)\right.\nonumber\\
\left.(1+\frac{\Omega_{\Lambda,0}}{70})\right]^{-1}.
\label{eq:g}
\end{eqnarray}   

The above formulae were given in Peacock \& Dodds (1996) and Peacock
(1997) for $z=0$. However Moscardini et al. (1997) argue that they
all apply to any cosmic epoch $z$, as long as one replaces $g$ by $g(z)$,
and interpret the quantity $x$ as the linear power-spectrum at epoch $z$:
\begin{eqnarray}
x=\Delta^2_{\rm{lin}}(k_0,z)=
\Delta^2_{\rm{lin}}(k_0)(1+z)^{-2}\left[g(z)/g(0)\right]^2.
\end {eqnarray}
Replacing $g$ by $g(z)$ implies writing both $\Omega$ and
$\Omega_{\Lambda}$ as a function of redshift. Their expressions have
been taken from Lahav et al. (1991) and obviously depend on the
underlying cosmology:  
\begin{eqnarray}
\Omega(z)=\Omega_0\;(h/h_0)^{-2}\;(1+z)^3\nonumber\\
\Omega_{\Lambda}(z)= \frac{\Lambda c^2}{3 h^2},\;\;
\;\;\;\;\;\;\;\;\;\;\;\;\;\;\;\;\;\;\;\;\;
\end{eqnarray}
where $\Lambda$ is the cosmological constant which does not vary with
time, and
\begin{eqnarray}
h(z)=h_0\left[\Omega_0(1+z)^3-(\Omega_0+\Omega_{\Lambda,0}-1)(1+z)^2+
\Omega_{\Lambda,0}\right]^{1/2}
\end{eqnarray}
\\
\noindent
Now that we have all the theory in place, we can finally evaluate the
evolution of 
the spatial two-point correlation function $\xi(r,z)$ with
time. Note that $\xi(r,z)$ is related to the power-spectrum $\Delta^2(k,z)$ via
the expression:
\begin{eqnarray}
\xi(r,z)=\int \Delta^2(k,z)\frac{\sin kr}{kr}\frac{dk}{k}
\label{eq:xi}
\end{eqnarray}
In order to compare theoretical models with data, $\xi(\bar{r},z)$ has
been obtained for all the redshifts sampled by different surveys. The
fiducial length $\bar{r}$ has been chosen to be $r_{max}/2$, where
$r_{max}$ is the upper limit of the range of scales used to measure
the clustering signal within each survey. This choice, even though
somewhat ``ad hoc'', seems plausible given that the measurements
go from about $r=0$ to $r_{max}$, 
and this gives an effective radius $\sim r_{max}/2$. 
A more exact analysis
would have to take into account the actual bin steps, weighting and
fitting procedure used in deriving $\xi$ for each survey, but since we 
were only interested in the relative scalings, we kept
it to this simple approximation. Note that the values of
$r_{max}$ (and therefore of $\bar{r}$) vary both with redshift and
cosmology and correspond to scales that range from linear to highly
non-linear (see Figure 2). 
In the cases where the 3-d clustering was obtained
by deprojecting the angular (2-d) correlation function $w(\theta)$ (as 
for instance in the HDF points), the value of 
$r_{max}$ was obtained via $r=\theta_{max} x$, with $\theta_{max}$ 
maximum angular scale and $x$ comoving
coordinate, whose expressions for different cosmologies 
are given in Appendix A.
We repeated this  analysis for the three cosmologies  introduced
in section 2.

\begin{figure*}
\vspace{16cm}  
\includegraphics{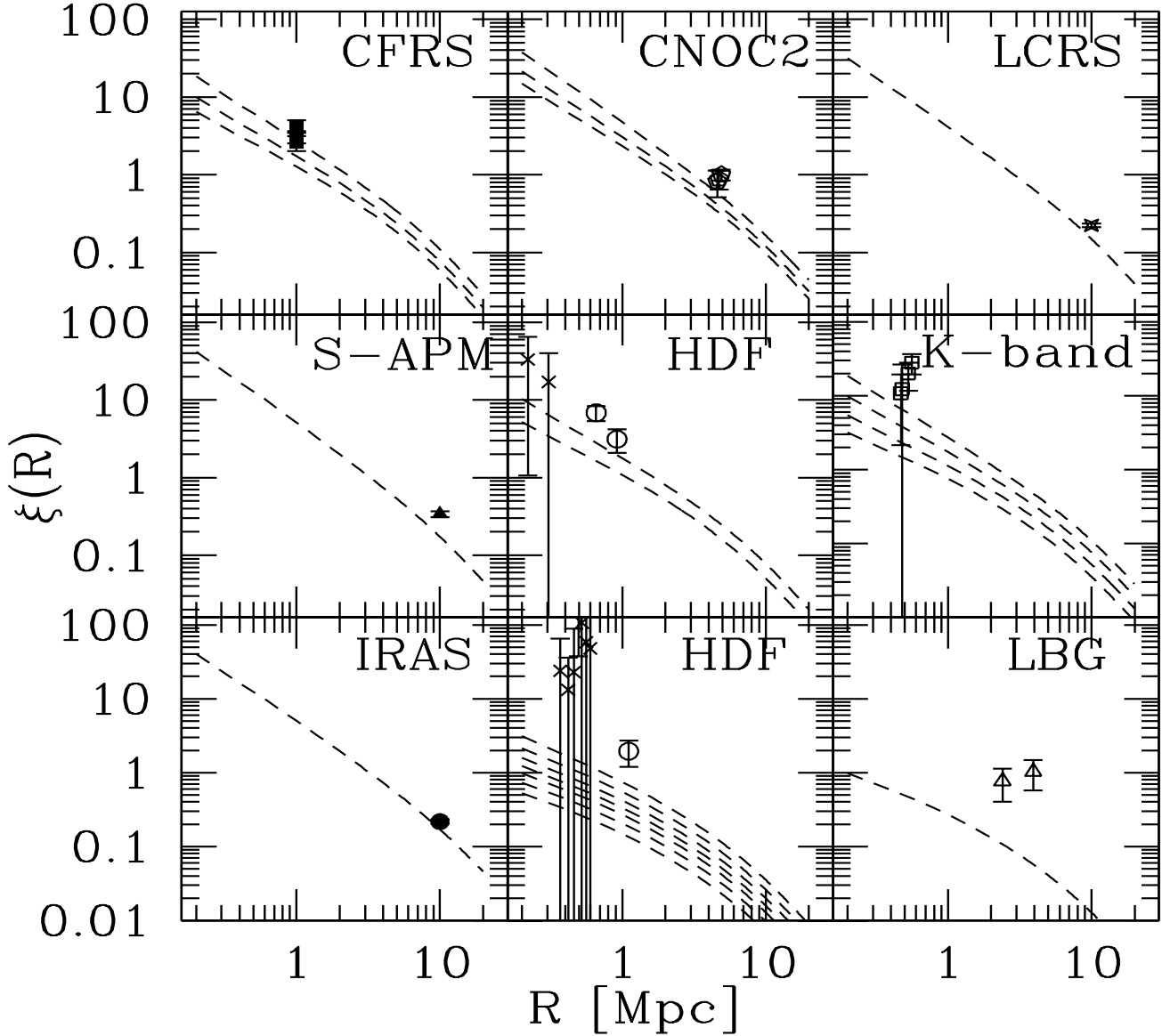}
\caption{Theoretical predictions for the spatial two-point correlation
  function $\xi(R)$ at different redshifts and for an Einstein-de
  Sitter Universe. Each panel corresponds to a
  different survey and the dashed lines indicate the models obtained for
  those redshifts $\bar{z}$'s which have available measurements of
  $r_0(\bar{z})$ (or alternatively of $\sigma_8(\bar{z})$, see table
  1). Lower lines correspond higher redshifts. 
The points show the values for $\xi(R)$
  obtained from the data in the assumption of a power-law form for the
  correlation function, where $R$ is half the size of the
  maximum scale used to derive the clustering measurements,
  representing the scale used to determine $r_0$. 
Different symbols correspond to different surveys. In the
  HDF panels crosses indicate the results from Magliocchetti \& Maddox
  (1999) (HDF$_2$), while empty circles are for Connolly, Szalay \&
  Brummer (1998) (HDF$_1$). In the LBG panel the higher point is taken
  from Adelberger et al. (1998).
\label{fig:xi1} }
\end{figure*} 

Figure 2 shows the theoretical predictions for the spatial two-point
correlation function $\xi(\bar{r}, z)$ at different redshifts and
compares them with the respective data.  Each panel corresponds to a
different survey (an exception has been made for the HDF in order not
to overcrowd the relative panel) and the different curves show the
predictions for $\xi$ at the redshifts $\bar z$ quoted for each of their
clustering measurements; the lower curves are for higher $z$'s.

The points show $\xi(\bar{r},z)$ derived from the observations
assuming a power-law form
$\xi(\bar{r},z)=\left[\bar{r}/r_0(z)\right]^{-\gamma}$.  
The values of $\gamma$ are directly observed, as listed in table 1 and
$r_0(z)$ is derived for each cosmology from expressions
(\ref{eqn:rr}) and (\ref{eqn:sigma}).
Note that the assumption of a power-law form for $\xi$ is justified by
the small areas covered by most of the surveys.
In the case of wide-area surveys (such as Stromlo-APM and IRAS)
$\bar{r}$ has been fixed to a standard value of 10 Mpc, independent of
cosmology.
Different symbols correspond to different surveys. In the
HDF panels crosses indicate the results from Magliocchetti \& Maddox
(1999) (HDF$_2$), while empty circles are for Connolly, Szalay \&
Brummer (1998) (HDF$_1$). In the LBG panel the higher point is taken
from Adelberger et al. (1998).
Figure 2 shows the great spread in the scales  sampled by different
surveys ranging from a few tenths to $\simgt$10 Mpc. 
It is also possible to note that even within the
same survey (e.g. HDF and LBG), different analysing techniques 
measure  different scales.  As we will show later in this
section, this scale effect can partially explain the apparent
discrepancy amongst results quoted for similar sets of data. 

\subsection{Populations}

Different surveys in general sample different populations of objects;
this is due to both the selection criteria and the redshift range
sampled by the survey. 
Objects selected in the UV band will be dominated by star-forming
galaxies, while for instance, B band selected objects will give mix of
early and late-type galaxies.
At higher redshifts the population mix sampled by a survey will also
depend on $z$ since the rest-frame pass band is shifted towards the
blue. 
For example a survey which selects objects in the I-band will contain
many early-type galaxies at low $z$, but for $z\simgt 1.5$ the
observed I-band corresponds to the rest-frame UV band, and so the
sample will be dominated by star-forming galaxies.
This also implies that, even within the same survey, the galaxy
population that is sampled depends strongly on the redshift.

Since late-type/star-forming galaxies cluster more weakly than
early-type ones (e.g Loveday et al. 1995; Hermit et al. 1996; Guzzo et
al. 1997), the changes in the galaxy population must be taken into
account when comparing the clustering measurements from different
surveys and different redshifts.

As a simple first step to minimizing this effect we  divided our sample
into different populations, according to the rest-frame selection
band. The Stromlo-APM survey samples galaxies in the rest  B-band (see
table 1) as it does the Las Campanas redshift survey (LCRS, objects selected
in the observed R-band at a median redshift $\bar{z}=0.13$) and the
CNOC2. The same holds for both the CFRS survey (I-band selected objects
at $0.3\simlt z \simlt 1$) and the HDF for $z\simlt 1.4$. We will
denote the objects selected by these surveys with the collective name
of {\it blue}. At higher redshift the observed I-band shifts into the
rest-frame UV band so that, for $z\simgt 1.4$, the population sampled
by the HDF will be dominated by star-forming galaxies. 
The selection criteria of both IRAS and Lyman-Break galaxies also lead
to samples dominated by galaxies undergoing star-formation.
We will call these objects {\it star-forming}. 
The Keck K-band survey selects objects in the near-IR, which for
$0.3\simlt z\simlt 1.4$ corresponds to a rest-frame R-band. We
therefore expect the populations sampled to be mainly 
early-type ({\it red}) galaxies.

Now that we have made this division into different populations according
to their star-forming activity and therefore their colours, we can go
back to the issues of the evolution of galaxy clustering and 
the redshift evolution of bias. In more detail, for each of the
three populations we have then evaluated the quantity $b(\bar{r},
z)=\sqrt{\left[\xi_g(\bar{r},z)/\xi_m(\bar{r},z)\right]}$, as
explained in section 4.1.

\begin{figure*}
\vspace{16cm}  
\includegraphics{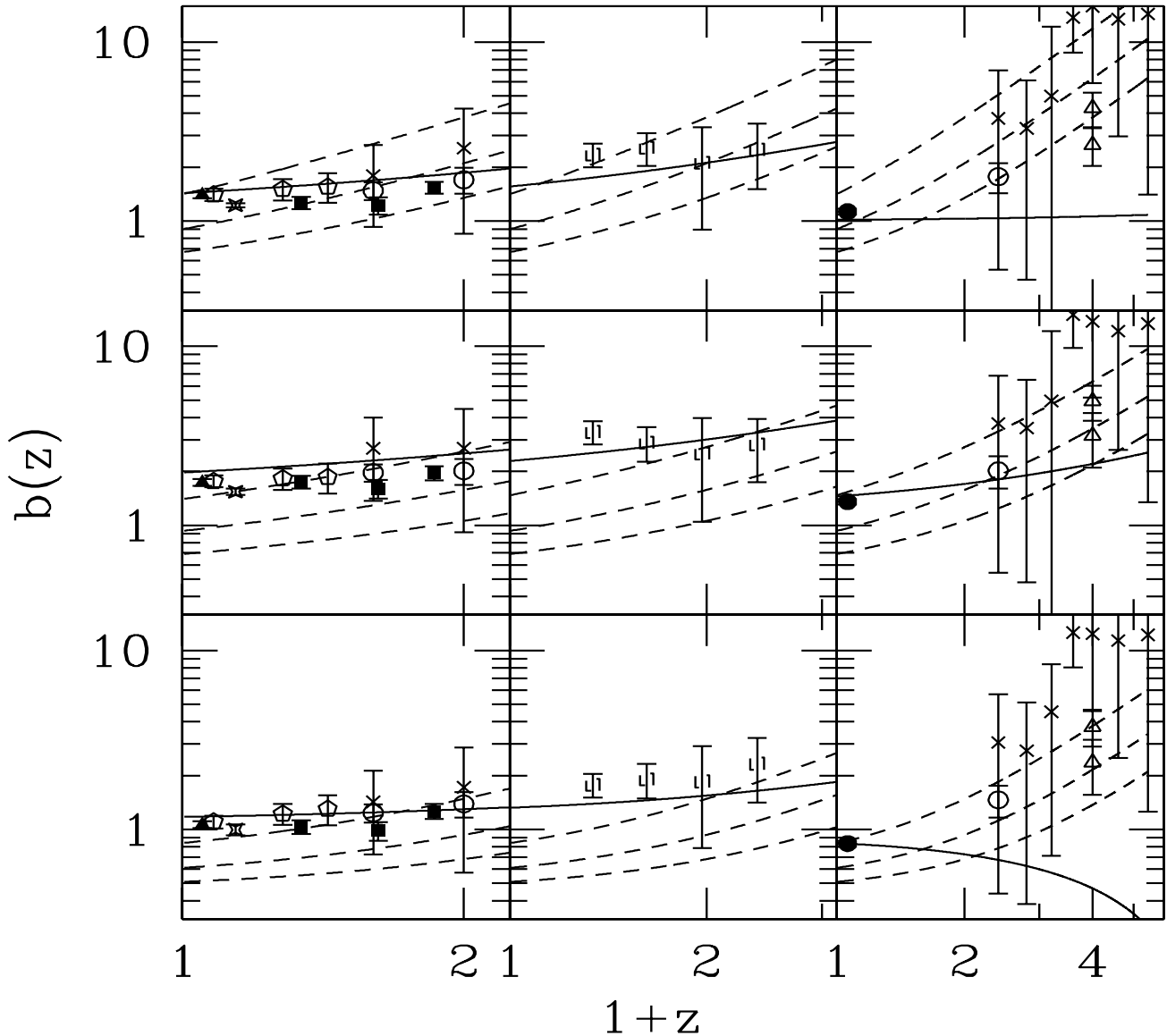}
\caption{Trend of the bias $b(z)$ as a function of redshift for
  different populations of objects. Left panels correspond to blue
  objects, middle panels to red objects and right panels to
  star-forming galaxies. Symbols are the same
  as in figure 2. Results have been obtained for an Einstein-de Sitter
  Universe (top), open Universe (centre) and flat Universe with a
  cosmological constant (bottom). Lines indicate theoretical models
  for the evolution of bias as explained in section 2. 
Solid lines represent the test-particle
  (B1) model, while the dashed lines are for the merging (B2) model
with $M_{min}= 10^{11}, 10^{12}$ and $10^{13}M_\odot$, 
  lower curves corresponding to lower halo-masses (see text for details).
\label{fig:b} }
\end{figure*} 

Figure 3 shows the trend of $b(z)$ as a function of $(1+z)$,
respectively for blue (left panels), red (middle panels) and
star-forming galaxies (right panels). Different symbols correspond to
different surveys and are the same as in figure 2. Once again 
the upper LBG point (empty triangle) is taken from Adelberger et
al. (1998). The upper panels are for an Einstein-de Sitter Universe
($\Omega_0=1$, $h_0=1$), the middle panels for open cosmologies
($\Omega_0=0.4$, $h_0=0.65$) and the lower panels correspond to flat
geometries with non-null cosmological constant ($\Omega_0=0.4$,
$\Omega_{\Lambda, 0}=0.6$, $h_0=0.65$).

There are many interesting features to note in Figure 3. The first is 
the trend of points at low
redshift ($z\simlt 1.5$). The large spread amongst the different
surveys seen in Figure 1 is much reduced after renormalizing the
amplitude according to the scale sampled by each survey. Panels on the
left-hand side of Figure 3 show the corrected results in excellent
agreement with each other. This good agreement also suggests that
there is little dependence of bias on scale.

The systematic higher clustering amplitude in the Keck K-band survey
as compared to other surveys (see middle panels in Figure 3) can now
be explained as due to different clustering properties of different
populations. In more detail the results show, as expected from local
measurements, that early-type objects cluster more strongly than
late-type ones.

In Figure~3 we have also plotted some theoretical predictions for the
evolution of bias with redshift.
The solid lines correspond to the test-particle model (B1), described
by equation (\ref{eqn:b1}).
Note that the value of the quantity $b_0=\sigma_{8,g}/\sigma_{8,m}$
(where in this case $\sigma_{8,m}$ is the non-linear \rm{r.m.s.} mass
density fluctuation at 8h$^{-1}$ Mpc, as calculated from the models
described in section 4.2) varies for different populations.
For the models we use a different value of $\sigma_{8,g}$ for each
galaxy type, with star-bursting galaxies being less clustered
than quiescent galaxies with old stellar populations.  
We estimate the amplitudes $\sigma_{8,g}$ from the values of $r_0$
measured for three subsamples from Stromlo-APM survey (Loveday, Tresse
\& Maddox, 1999): for galaxies with no emission lines (red objects)
$\sigma_{8,g}=1.13$; galaxies with weak emission (blue galaxies)
$\sigma_{8,g}=0.93$; and $\sigma_{8,g}=0.66$ for galaxies with strong
emission lines (star-forming galaxies).

The dashed lines correspond to the merging model for the evolution of
bias with $z$ (B2), with lower curves corresponding to lower
halo-masses ($10^{11}M_{\odot} \le M_{min} \le
10^{13}M_{\odot}$, see section 2). Note that neither the B1 nor the B2
curves have been anchored to any $z=0$ point.

At low redshift, quite independent of the cosmology, the B1 models
seem to provide a reasonably good fit to the data, both in the case of
blue and early-type objects, even though we cannot rule out the
constant bias model as a possible description (B0 - equation
\ref{eqn:b0}, lines parallel to the $x$ axis, not shown in the figure).
This is also seen in the analysis of the clustering of radio galaxies
(e.g. Magliocchetti et al. 1999); in this case the mean redshift for
the clustering measurements is about $z\sim 1$.
Even though the data cannot distinguish between constant bias (B0)
models and models with bias linearly evolving with redshift (B1), it
is clear that the clustering measurements obtained for the radio
sample in such a redshift range are in conflict with the predictions
obtained from the merging (B2) model.

At higher redshifts ($z\simgt 2$) though, both the constant bias and
the test particle model grossly fail to describe the data which show
very high values for $b(z)$. 
Note that in this redshift range we have measurements only for
star-forming objects, therefore it is impossible to state whether this
rise in the level of biasing is always true for $z\simgt 2$,
independent of the population, or only holds for star-forming
galaxies.
Although the match is not perfect,
the merging model correctly describes the trend of the data (and also
matches the IRAS points), with lower halo-masses 
($M_{min}=10^{12}M_\odot$) required for an Einstein-de Sitter Universe
and higher halo-masses ($M_{min}=10^{13}M_\odot$) for low-density
models (independent of $\Omega_{\Lambda,0}$).

\subsection{Malmquist bias}

Now we turn to a discussion of the selection effects, namely the
Malmquist bias and K-correction.
 
Although it is not possible to correct for the Malmquist-bias and
K-correction, models can be made to allow for their effects.  In
a detailed model (e.g. by using the semi-analytic approach;
\cite{garching,durham,so}), these selection effects can be convolved
with the model and direct comparison can be made with the
observations.  However, these models contain a large number of
parameters and the gain in details is accompanied by a loss of
intuitive understanding.  In order to keep the discussion simple and
independent of model parameters, we will not use this approach here.  

Instead, by dividing the objects into different populations we
minimize the K-correction problem, because objects belonging to the
same population are likely to have similar spectra, and so have
similar K-corrections.

As already mentioned in the introduction, more luminous galaxies 
found to cluster more strongly \cite{park_cfa2,Lo} than fainter ones.
Therefore, for the apparent magnitude selected samples in Figure 3 we
would expect to see an increase of the clustering introduced by the
Malmquist bias at higher $z$'s.
The CFRS measurements are based on all galaxies brighter than the
apparent magnitude limit of the survey, and so the points sample
objects with increasing absolute magnitudes as one goes to higher
redshifts. 
We can see a hint of  the expected  Malmquist effect in these points
in that the third bin (centred at $z\simeq 0.87$) shows a
slight increase in clustering amplitude, and includes only galaxies
brighter than $L_B^{*}$ (see table 1). 
However this is at a very low significance level. 
On the other hand the measurements from the CNOC2 and the Las Campanas 
surveys are based
on sub-samples of galaxies with a cut-off in absolute magnitude
(respectively $L_R\ge -20$ and $-23 \le L_R\le -18$, and so should not
show this effect.

In the Keck K-band survey (panels in the centre of Figure 3),
clustering measurements have been obtained for all the galaxies
respectively with $L_K\ge -23$ in the three higher redshift bins and
with $L_K\ge -21.5$ in the lower redshift range. We therefore expect
the lower-$z$ point to underestimate the clustering with respect to
the results for higher $z$'s and brighter absolute
magnitudes. However, given that $L_K^{*}\sim -25+5\;\rm{log}(h_{50})$,
even for $L_K\ge-23.5$, all the galaxies still sit on the flat
faint-end of the luminosity function. Since  Malmquist bias is likely to
be important only for luminosities $\simgt L^*$, we argue that the effect
should small in this case.

Unfortunately we do not have reliable information on the absolute
luminosities of objects in the HDF because the K-corrections
are highly uncertain, particularly for $z\simgt 1$. 
However, at low redshift ($z\simlt 1.4$), the faint apparent magnitude
limit suggests that all galaxies will be below $L^*$, and so Malmquist
bias should not play an important role. This is consistent with the
trend shown by the data in Figure 3 (left panels).  At higher redshift
the galaxies in the sample are  brighter than $L^*$, therefore we might expect
Malmquist bias to be more important, so that the clustering amplitude
would increase with increasing redshift as seen in the data.

It would be possible to quantify the Malmquist bias effect by 
converting the absolute
luminosity limit for each subset of galaxies to an equivalent mass
limit and hence find the expected clustering amplitude for each
redshift.  However the K-corrections, and hence the derived absolute
magnitudes, are highly uncertain at these redshifts.
Also the mass-to-light ratio is unknown at high redshift, and given
the rapid evolution of stellar populations at $z<1$, it is likely to
be very different to local galaxies.
An alternative way to tackle this problem is to consider the space
density of objects. 
In models based on high peak biasing, rarer objects correspond to
higher masses, which are more strongly clustered. Hence a comparison 
between the observed and predicted comoving space-density of galaxies
allows us to estimate the effective $M_{min}$ for each redshift bin,
independent of uncertainties in the absolute magnitudes. 
This should provide a more robust approach to quantifying the
Malmquist bias.

\begin{figure}
\vspace{8.5cm}  
\includegraphics{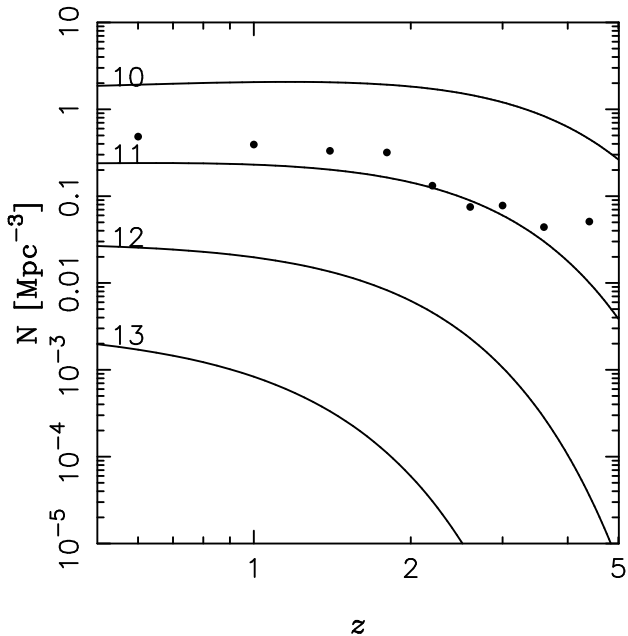}

\caption{The estimated comoving galaxy space density as a function of
redshift. The points show the estimates for the HDF galaxies as used
in the clustering measurements. The lines show the Press-Schechter
prediction for the density of halos with mass greater than
$M_{min}=10^{10}, 10^{11}, 10^{12}$ and $10^{13}M_\odot$, as labelled.
We applied the standard Press-Schechter formalism using the cosmology and
power spectrum from our model I.
The data at $z\simgt 3$ are almost certainly affected by random
errors, because the space density implies a low $M_{min}$, whereas the
clustering implies a very high $M_{min}$ (see Figure~\ref{fig:b}).
\label{fig:rhoz} }
\end{figure} 

In Figure~\ref{fig:rhoz} the points show the observed mean density for
each HDF redshift bin, estimated as $\bar{N}=N_{\Delta z}/V$, where $V$ is
the comoving volume in each bin assuming an Einstein-de Sitter Universe. 
Note that, as expected, there is a clear trend for the space-density
to decrease with increasing redshift, although the decrease is not
very strong, even at high redshift. 
The lines show the space density of 
haloes of mass greater than $M_{min}=10^{10}, 10^{11}, 10^{12}$ and
$10^{13}M_\odot$ predicted from the standard Press-Schechter formalism
with the power spectrum as in our model I. 
For $z\simlt 2$ the points are consistent with a fixed halo
mass $\sim 5 \times 10^{10} M_\odot$. 
For $2\simlt z \simlt 4$ there is a hint that $M_{min}$ may rise to
$\sim 10^{11} M_\odot$, but the uncertainties are large, and for
$z\simgt 4$ the apparent decrease is not significant. 
The merging model (B2) introduced by Matarrese et al. (1997) and
Moscardini et al. (1998) predicts the evolution of bias with redshift,
according to the mass of the halos.
Thus we can compare directly to the model (B2) (dashed lines in the
panels on the right-hand side of Figure 3) with a minimum halo mass
$M_{min}(z)$ which is roughly constant or slightly increasing with
at higher redshifts (see also Arnouts et al. 1999).


\section{CONSTRAINTS ON $\Omega_0$}

The rate of evolution of the galaxy correlation function with redshift
has been used to estimate Cosmological parameters \cite{peacock_egc}.
This approach assumes that the bias does not have any scale
dependence, and, it does not vary significantly in time.
However, recent studies have questioned this approach and have shown
that the time evolution of bias is very important, even if we can
choose to ignore its scale dependence at large scales.  The most
remarkable manifestation of the time evolution of bias are the Lyman
break galaxies, that have a clustering amplitude comparable to present
day galaxies.  The models of bias evolution discussed in section 2 
also suggest that the time evolution of bias is very important.  

In principle, it is possible to use the models of bias along with the 
observations to see if the observed redshift evolution of clustering
is consistent with a given model or not.  However, this approach
requires the power spectrum for density fluctuations as an input, and
hence is model dependent.  

We propose a simple test to derive a lower limit on $\Omega_0$.  We
make use of the fact that in all realistic scenarios, bias 
always increases as we go to higher redshifts.  This monotonic
increase in bias implies that the galaxy correlation function will
evolve at a slower rate than the correlation function of dark matter,
i.e. $\dot{D}_{gal}(z) \leq \dot{D}_{m}(z)$.  $D_{m}(z)$ depends
only on $\Omega_0$ (and $\Omega_{\Lambda,0}$), and evolves at a slower
rate for lower $\Omega_0$.  If we compute the allowed values of
$\dot{D}_{gal}(z)$ in a given survey, and we get a firm lower limit on
this rate, then we can rule out those cosmologies that predict
$\dot{D}_{m}(z) < \dot{D}_{gal, obsv}(z)$.  {\it We would like to
stress, that this method is completely model independent if used at
large/linear scales.}

\begin{figure}
\epsfxsize=3.3truein\epsfbox[17 150 580 700]{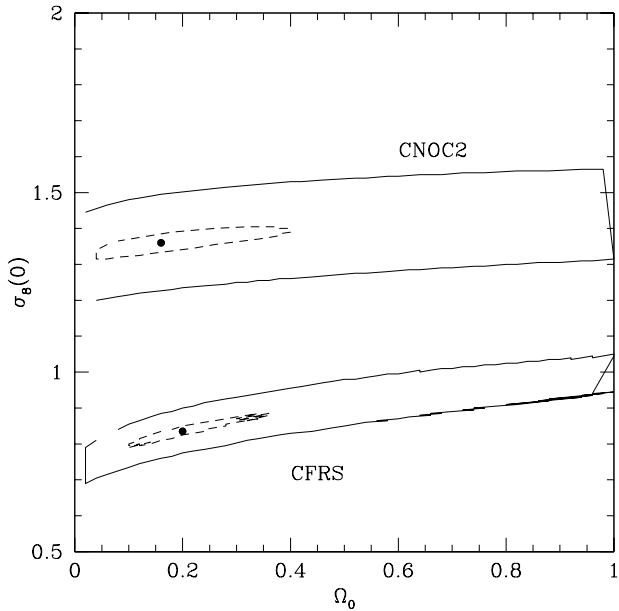}
\caption{This figure shows the best fit values of $\Omega_0$ and
$\sigma_{8,g}(0)$ for CFRS and CNOC2 surveys assuming an open universe
with $\Omega_{\Lambda,0}=0$.  The $68\%$ confidence
limits are shown as thick lines.  Dashed lines show the same if error
bars are reduced to $1/5$ of their present size.  Allowed values for
$\Omega_0$ lie on the right-hand side of the diagram. See text for
details.\label{fig:jas}}
\end{figure}

To further illustrate this idea, we have plotted the best fit points
for CFRS and CNOC2 surveys on the $\Omega_0$--$\sigma_8$ plane in
figure~\ref{fig:jas}.  Thick lines show the confidence limits at the
$68\%$ level.  The error bars at this confidence level are too big to be
of much use.  We have not shown data from any other survey listed in
this paper because the error bars for those are even larger.  However,
to demonstrate what can be achieved by the next generation surveys,
like 2DF (Maddox 1998), Sloan (Kim et al. 1999) and
VIRMOS (Garilli et al. 1999) which will have many more galaxy redshifts, we
have shown the contours for error bars reduced by a factor 5.  In this
case, one may begin to rule out interesting regions of the parameter
space.  

This test is more effective at lower redshifts as the difference in
the growth rate in different cosmologies is more striking.  Thus
surveys like Sloan and 2DF may provide some useful constraints through
this test.

\section{CONCLUSIONS}

We have presented a detailed analysis of the evolution of galaxy
clustering with redshift, specifically aimed at constraining 
epoch-dependent models for the bias.

As a first step we put together all the available measurements 
from different surveys, covering $z\simeq 0$ to $z\simeq 4.5$, and we
calculated the values for the \rm{r.m.s} galaxy fluctuation density on
a 8$h^{-1}$ Mpc scale, $\sigma_{8,g}$. This has been repeated for
different cosmologies. 
We find that the amplitude of galaxy clustering shows a general trend
of decreasing at redshifts $z < 2$ and increasing beyond that.

The amplitudes for different surveys show a remarkable scatter between
each other, and there are several factors which might be responsible
for this. First is the ``scale-dependence'' problem, coming from the
fact that the various surveys measure clustering at different scales.
Second is that different surveys select different populations of
objects which do not have similar clustering amplitudes.  Third is the
Malmquist bias, since brighter objects cluster more strongly than
fainter ones.
By using the Peacock \& Dodds (1994, 1996)
models for non-linear evolution of the power-spectrum $P(k,z)$ (and
consequently of the two-point correlation function $\xi(r,z)$), we
corrected for the scale-dependence effect and obtained values for the
bias written as $b^2(\bar{r},z)=\xi_g(\bar{r},z)/\xi_m(\bar{r},z)$,
where $\bar{r}$ is half of the maximum scale used within each survey
to work out estimates on the clustering amplitude. Note that this
approach can in principle also show a dependence of bias on scale.
We allowed for the population selection effect by dividing the objects
into three categories, according to the rest-frame pass band of their
selection  (roughly corresponding to their star-forming activity) and 
analyzing the evolution of clustering for each of these populations
separately. 
We also considered the expected variations introduced by the Malmquist
bias.  

We find that the main cause for the scatter amongst 
different surveys is the scale-dependence
problem. Renormalizing the amplitudes according to the physical scale
at which each measurement was taken brings the results in excellent
agreement with each other.

The main conclusion of this work is that the bias grows monotonically
from the present epoch to high redshifts, and the rate at which bias
grows increases rapidly as we go towards higher redshifts.

At low redshift the trend of the data-points seems to suggest a bias
whose functional form is a weakly increasing function of the
redshift. The test-particle bias is a useful model that
allows us to predict the evolution of bias for objects that are
positively biased. 
We do not see strong evidence for scale dependent bias or Malmquist
effects, but  the uncertainties are large. 

At higher redshifts the clustering signal appears to come from objects
which are highly biased with respect to the underlying distribution of
mass. The merging model for the evolution of bias (B2) can correctly
describe the steep rise in clustering amplitude seen for $z\simgt
2$. We note however that there are several uncertainties in the
interpretation of clustering at such redshifts. In particular, at high
$z$ we have measurements only for the population of star-forming
galaxies, and the lack of reliable absolute magnitude estimates for
these objects makes it impossible to apply any quantitative correction
for the Malmquist bias.

Lastly we find that the rate at which the observed amplitude decreases
at low redshifts is 
slower than the linear rate of evolution for density perturbations in
dark matter in most models.  We present arguments which allow us to
rule out all models for which this is not the case, i.e. the rate of
evolution of galaxy clustering is faster than the linear rate.
Present observations do not rule out any model but future observations
will allow us to constrain the density parameter.  We would like to
emphasise that this test does not depend on any detailed modelling of
galaxies/halos and so should provide a reliable constraint.

\vspace{0.3cm}
\noindent
\section*{ACKNOWLEDGMENTS}

MM acknowledges support from the Isaac Newton Scholarship. Jarle
Brinchmann and George Efstathiou are warmly thanked for very helpful and
stimulating discussions. We also thank the anonymous referee for
comments which improved the content of the paper.

\appendix
\section{Useful Cosmological Quantities}
We express here the quantities appearing in equation \ref
{eqn:rr}. Note that $P(\Omega_0, z)$ is obtained by direct integration
of the Limber equation (Peebles, 1980).\\

\noindent
\bf{A1 \hspace*{1cm} Case $\Omega_{\Lambda,0}=0$}
\begin{eqnarray}
P(\Omega_0,z)=\frac{\Omega_0^2(1+z)^2(1+\Omega_0
  z)^{1/2}}{4(\Omega_0-1)[(1+
\Omega_0z)^{1/2}-1]+\Omega_0^2(1-z)+2\Omega_0 z},
\label {eqn:P1}
\end{eqnarray}
\begin{eqnarray}
F(x)=\left[1-\left(\frac{H_0 x}{c}\right)^2(\Omega_0 -1)\right]^{1/2}
\label{eqn:F1}
\end {eqnarray}
\begin{eqnarray}
x=\frac{2c}{H_0}\left[\frac{\Omega_0 z-(\Omega_0-2)(1-\sqrt{1+\Omega_0
      z})}{\Omega^2_0 (1+z)}\right],
\label{eqn:x1}
\end{eqnarray}
\\

\noindent
\bf{A2 \hspace*{1cm} Case $\Omega_0+\Omega_{\Lambda,0}=1$}
\begin{eqnarray}
P(\Omega_0,z)=\Omega_0^{1/2}[(1+z)^3+\Omega_0^{-1}-1]^{1/2},
\label {eqn:P2}
\end{eqnarray}
\begin{eqnarray}
F(x)=1
\end {eqnarray}
\begin{eqnarray}
x=\frac{c}{H_0}\Omega_0^{-1/2}\int_0^z\frac{dz}{\left[(1+z)^3+
\Omega_0^{-1}-1\right]^{1/2}},
\label {eqn:x2}
\end{eqnarray}

\label{lastpage}


\begin{thebibliography}{}

\bibitem[\protect\citename{Adelberger et al.} 1998]{Ad} Adelberger
K.L., Steidel C.C., Giavalisco M., Dickinson M., Pettini M., Kellog
M., 1998, ApJ, 505, 18

\bibitem[\protect\citename{Arnouts et al.} 1999]{Ar}
Arnouts S., Cristiani S., Moscardini L., Matarrese S., Lucchin F.,
Fontana A., Giallongo E., 1999, astro-ph/9902290

\bibitem[\protect\citename{Bagla} 1998]{Bag} Bagla J.S., 1998, MNRAS,
299, 417

\bibitem[\protect\citename{Bardeen} 1986]{Bar} 
Bardeen J.M.,Bond J.R., kaiser N., Szlalay A.S., 1986, ApJ, 304, 15

\bibitem[\protect\citename{Baugh et al.} 1999]{durham} Baugh C.M.,
Benson A.J., Cole S., Frenk C.S., Lacey C.G., 1999, MNRAS, 305, 21

\bibitem[\protect\citename{Blanton et al.} 1998]{bias_phy} Blanton M.,
Cen R., Ostriker J.P., Strauss M.A., 1998, astro-ph/9807029

\bibitem[\protect\citename{Brainerd and Villumsen} 1994]{tgb_corr}
Brainerd T.G. and Villumsen J.V., 1994, ApJ, 431, 477

\bibitem[\protect\citename{Bunn and White} 1997]{Bunn} Bunn E.F.,
White M., 1997, ApJ, 480, 6

\bibitem[\protect\citename{Carlberg et al.} 1997]{Car} Carlberg R.G.,
Cowie L.L., Songaila A., Hu E.M., 1997, ApJ, 484, 538

\bibitem[\protect\citename{Carlberg et al.} 1998]{Car1} Carlberg R.G.,
Yee H.K.C., Morris S.L., Lin H., Sawicki M., Wirth G., Patton D.,
Shepherd C.W., Ellingson E., Schade D., Pritchet C.J., Hartwick
F.D.A., 1998, astro-ph/9805131

\bibitem[\protect\citename{Carrol et al.} 1992]{Carr}
Carrol S.M., Press W.H., Turner E.L., 1992, ARA\&A, 30, 499

\bibitem[\protect\citename{Col\'in et al.} 1998]{pedro} Col\'in P.,
Klypin A.A., Kravstov A.V., Kokhlov A.M., 1998, astro-ph/9809202

\bibitem[\protect\citename{Connolly, Szalay and Brummer} 1998]{Co}
Connolly A.J., Szalay A.S., Brummer R.J., 1998, ApJ, 499, L125

\bibitem[\protect\citename{Dekel and Lahav} 1998]{DL}
Dekel, A., \& Lahav, O., 1998,  ApJ, submitted (astro-ph/9806193)

\bibitem[\protect\citename{Fernadez-Soto et al.} 1998]{soto}
Fernadez-Soto A., Lanzetta K.M., Yahil A., 1999, ApJ, 513, 34

\bibitem[\protect\citename{Fry} 1996]{Fry} Fry J.N., 1996, ApJ, 461,
L65

\bibitem[\protect\citename{garilli et al} 1999]{gar}
Garilli B., Bottini D., Tresse L., LeFevre O., Saisse M., Vettolani
G., 1999, astro-ph/9907320

\bibitem[\protect\citename{Giavalisco et al.} 1998]{Gia} Giavalisco
M., Adelberger K.L., Steidel C.C., Dickinson M., Pettini M.,
Kellog M., 1998, ApJ, 503, 543

\bibitem[\protect\citename{Guzzo et al.} 1997]{guzzo} Guzzo L.,
Strauss M.A., Fisher K.B., Giavanelli R., Haynes M.P., 1997, ApJ, 489,
37

\bibitem[\protect\citename{Hamilton et al.} 1991]{ham}
Hamilton A.J.S., Kumar P., Lu E., Mathews A., 1991, ApJ, 374, L1


\bibitem[\protect\citename{Hermit et al.} 1996]{ors_corr} Hermit S.,
Santiago B.X., Lahav O., Strauss M.A., Davis M., Dressler A, Huchra
J.P., 1996, MNRAS, 283, 709

\bibitem[\protect\citename{Huan et al.} 1996]{lascamp} Huan L.,
Kirshner R.P., Shectman S.A., Landy S.D., Oemler A., Tucker D.L.,
Schechter P.L., 1996, ApJ, 471, 617

\bibitem [\protect\citename{Jenkins et al.} 1998]{Jen} Jenkins A.,
     Frenk C.S., Pearce F.R., Thomas P.A., Colberg J.M., White S.D.M.,
Couchman H.M.P., Peacock J.A., Efstathiou G., Nelson A.H., 1998, ApJ,
     499, 20

\bibitem[\protect\citename{Kaiser} 1984]{kaiser} Kaiser N., 1984,
ApJL, 284, L9

\bibitem[\protect\citename{Kauffmann et al.} 1999]{garching} Kauffmann G.,
Colberg J.M., Diaferio A., White S.D.M., 1999, MNRAS, 303, 188

\bibitem [\protect\citename{Kim et al.} 1999]{Kim}
Kim R.S.J., Kepner J.V., Strauss M.A., Bahcall N., Gunn J.E., Lupton
R.H., Schlegel D.J., 1999, AAS, 194, 8801

\bibitem[\protect\citename{Le Fevre et al.} 1996]{Le} Le Fevre O.,
Hudon D., Lilly S.J., Crampton D.,  Hammer F., Tresse L., 1996,
ApJ, 461, 534

\bibitem[\protect\citename{Lahav et al.} 1991]{La}
Lahav O., Lilje P.B., Primack J.R., Rees M.J., 1991, MNRAS, 251, 128

\bibitem[\protect\citename{Loveday et al.} 1995]{Lo} Loveday J.,
Maddox S.J., Efstathiou G., Peterson B.A., 1995, ApJ, 442, 457

\bibitem[\protect\citename{Lovedays et al.} 1999]{Lo1} 
Loveday J., Tresse L., Maddox S.J., 1999, in preparation.

\bibitem[\protect\citename{Maddox} 1998]{mad}
Maddox S.J., 1998,  1998yugf.conf..198M

\bibitem[\protect\citename{Magliocchetti et al.} 1999]{Mag}
Magliocchetti M., Maddox S.J., Lahav O., Wall J.V., 1999,
MNRAS, 306, 943

\bibitem[\protect\citename{Magliocchetti and Maddox} 1999b]{Mag1}
Magliocchetti M., Maddox S.J., 1999, MNRAS, 306, 988

\bibitem[\protect\citename{Matarrese et al.} 1997]{Mat} Matarrese S.,
Coles P., Lucchin F., Moscardini L., 1997, MNRAS, 286, 115

\bibitem[\protect\citename{Mo and White} 1996]{Mo} Mo H., White
S.D.M., 1996, MNRAS, 282, 347

\bibitem[\protect\citename{Moscardini et al.} 1998]{Mos}
Moscardini L., Coles P., Lucchin F., Matarrese S., 1998, MNRAS,
299, 95

\bibitem[\protect\citename{Narayanan, Berlind and Weinberg} 1998]{Narayanan}
Narayanan V.K., Berlind A.A., Weinberg D.H., 1998, submitted to ApJ,
astro-ph/9812002

\bibitem[\protect\citename{Nusser and Davis} 1994]{Nu} Nusser A.,
Davis M., 1994, ApJ, 421, L1

\bibitem[\protect\citename{Park et al.} 1994]{park_cfa2} Park C.,
Vogeley M.S., Geller M.J., Huchra J.P., 1994, ApJ, 431, 569

\bibitem[\protect\citename{Pe1} 1994]{peacock1} 
Peacock J.A., Dodds S.J., 1994, MNRAS, 280, L19

\bibitem[\protect\citename{Peacock2} 1996]{peacock2} 
Peacock J.A., Dodds S.J., 1996, MNRAS, 267, 1020 

\bibitem[\protect\citename{Peacock} 1997]{peacock_egc} Peacock
J.A. 1997, MNRAS, 284, 885

\bibitem[\protect\citename{Peebles} 1980]{Pe} Peebles P.J.E., 1980,
{\it The Large-Scale Structure of the Universe}, Princeton University
Press

\bibitem[\protect\citename{Press and Schechter} 1974]{ps74} Press
W.H., Schechter P., 1974, ApJ, 187, 425

\bibitem[\protect\citename{Shepherd et al.} 1996]{She} Sheperd C.W.,
Carlberg R.G., Yee H.K.C., Ellingson E. 1997, ApJ, 479, 82

\bibitem[\protect\citename{Saunders et al.} 1992]{Sa} Saunders W.,
Rowan-Robinson M., Lawrence A., 1992, MNRAS, 258, 134

\bibitem[\protect\citename{Somerville \& Primack} 1998]{so}
Somerville R.S., Primack J.R., 1998, Astro-ph/9802268

\bibitem[\protect\citename{Su} 1995]{su}
Sugiyama N., 1995, ApJS, 100, 281

\bibitem[\protect\citename{Tegmark and Peebles} 1998]{teg} Tegmark
M. and Peebles P.J.E. 1998, ApJL 500, L79

\bibitem[\protect\citename{Tresse }1999]{Tresse}
Tresse, L., 1999, Les Houches Summer School 1999, astro-ph/9902209

\bibitem[\protect\citename{Treyer }1996]{Tre}
Treyer M.A., Lahav O., 1996
MNRAS, 280, 469

\end{thebibliography}
\end{document}